\documentclass[prl,a4paper,twocolumn,showpacs,superscriptaddress,floatfix]{revtex4}
\usepackage{amsmath,amsthm}
\usepackage{amsfonts}
\usepackage[hmargin=2cm,vmargin=2cm,bmargin=2cm]{geometry}
\usepackage{graphicx}

\theoremstyle{definition}

\theoremstyle{remark}

\usepackage[latin1]{inputenc}
\usepackage{graphicx}
\usepackage{braket}
\usepackage{pdfpages}
\usepackage{tensor}
\usepackage{subfigure}

\begin{document}
\title{Tunable ${\cal \chi/PT}$ Symmetry in Noisy Graphene}
\author{E. Frade Silva}
\affiliation{Departamento de F\'isica, Universidade Federal da Para\'iba, 58051-970, Jo\~ao Pessoa, Para\'iba, Brazil.}
\author{A. L. R. Barbosa}
\affiliation{Departamento de F\'isica, Universidade Federal Rural de Pernambuco, Dois Irm\~aos, 52171-900 Recife, Pernambuco, Brazil}
\author{M. S. Hussein}
\affiliation{Instituto de Estudos Avan\c cados and Instituto de F\'{\i}sica, Universidade de S\~{a}o Paulo, C.P.\ 66318, 05314-970 S\~{a}o Paulo, SP, Brazil and Departamento de F\'{i}sica, Instituto Tecnol\'{o}gico de Aeron\'{a}utica, CTA, S\~{a}o Jos\'{e} dos Campos, S.P., Brazil.}
\author{J. G. G. S. Ramos}
\affiliation{Departamento de F\'isica, Universidade Federal da Para\'iba, 58051-970, Jo\~ao Pessoa, Para\'iba, Brazil.}
\date{\today}
\begin{abstract}

We investigate the resonant regime of a mesoscopic cavity made of graphene or a doped beam splitter. Using Non-Hermitian Quantum Mechanics, we consider the Bender-Boettcher assumption that a system must obey parity and time reversal symmetry. Therefore, we describe such system by coupling chirality, parity and time reversal symmetries through the scattering matrix formalism and apply it in the shot noise functions, also derived here. Finally we show how to achieve the resonant regime only by setting properly the parameters concerning the chirality and the PT symmetry.
\end{abstract}
\maketitle

\section{I. Introduction}

Many current investigations propose Hermitian models to describe dissipative processes in order to maintain the Dirac's assumption and obtain real observable values\cite{ref32, ref33, ref34}. Although it has been demonstrated to be a sufficient condition to obtain acceptable numbers, it has never been proven that it is necessary. For instance, non-Hermitian Hamiltonians are treated as a standard technique to describe open systems with dissipative processes\cite{ref31,ref17}, and the most known example is the radioactive decay\cite{ref28}. Beyond that, there is an attempt of a possible generalization of quantum mechanics postulates for which the open systems may be described by parity and time reversal invariant Hamiltonians.

In general, Parity and Time Reversal Symmetry (PT) has been used as a physical artifice very useful to establish selection rules and to settle some properties of a quantum system. These features allowed one to predict fundamental particles; {\it e.g.} the kaons, and to engender philosophical debates in the very foundations of physics, as the time arrow problem and, concomitantly, the PT-symmetry description of quantum mechanics.

The latter was proposed by Bender and Boettcher\cite{ref30} when they analyzed a non-Hermitian Harmonic Oscillator Hamiltonian and figured out that its spectrum is entire real, unless the PT-symmetry is spontaneously broken. Ever since, a plenty of works in this matter were published, both theoretically and experimentally, indicating that might be possible to obtain acceptable quantum mechanics using such assumption. Important application of ${\cal PT}$ symmetry in optics were developed. In particular we mention two coupled ${\cal PT}$ symmetric waveguides: a ${\cal PT}$ symmetric Bragg scatterer and a ${\cal PT}$ microring laser \cite{ref39, ref40}

Moreover, fundamental symmetries, such as time reversal, spin rotation, chirality etc, must be considered to describe cavities with a large number of resonances ({\it e.g.} heavy nuclei). Accordingly, the treatment given to this complex systems is to analyze their symmetries, ignoring its detailed internal properties\cite{ref38}. The different combination of the absence, or presence, of these symmetries forms the tenfold Cartan classes\cite{ref11}. In this way, it turns out to be natural to study systems such as mesoscopic cavities in the view of Bender-Boettcher formalism.

In principle, for a complex system, the resulting Hamiltonian must be random, but here, we address a deterministic behavior to the system  for simplicity, which will allow the study of both quantum mechanical formalisms. We approach the problem by the scattering matrix theory, which seems the most simple way to couple the desired symmetries. In this paper, we restrict ourselves to the effect caused by the coupling of Chirality and/or PT-symmetry. We chose such couplings because, as will be seen throughout the manuscript, chirality plays a very important role in modern mesoscopic cavities.

Since the characteristic scattering matrices are obtained, we derive the shot-noise functions which act as observables to quantify the coupled symmetries. Then, we demonstrate how the resonance regime condition of such systems can be achieved and its dependence with parameters which govern the degree of symmetry. We further find corrections to the transport functions of a PT cavity tuning the relative chirality parameter.

\section{II. Scattering Matrix: Chirality, Parity and Time Reversal}
We begin by introducing the chiral symmetry and its applications to physics. For a two-dimensional bipartite lattice system, there is a two-fold degeneracy in the spectrum. It is said that the system supports a chiral symmetry and may be described by the off-diagonal Hamiltonian\cite{ref12}
\begin{equation}\label{eq46}
\mathcal{H}=\left(\begin{array}{cc}
0 & A\\
A^\dagger &0
\end{array}\right)
\end{equation}
The off-diagonal terms $A$ are called hopping matrices, since one considers the transition between the sublattices. Now we introduce the chiral operator, $\hat{C}$, which brings up a degeneracy in each energy spectrum level due to the bipartite lattice
\begin{eqnarray}\label{eq47}
&\left\{\mathcal{H}, \hat{C}\right\}=0,&\nonumber \\
&\mathcal{H} = -\hat{C} \mathcal{H} \hat{C}&.
\end{eqnarray}
Comparing (\ref{eq47}) with (\ref{eq46}), the chiral operator can be represented by the matricial form 
\begin{equation}\label{eq48}
\hat{C} = \left(\begin{array}{cc}
\mathbb{I}_N & 0 \\ 
   0     & -\mathbb{I}_N
\end{array}\right).
\end{equation}
These features about the chiral operator are used to explore the topological states of condensed matter systems, such as crystalline \cite{ref13} and Chern insulators\cite{ref14}. Here we investigate graphene \cite{ref15}, another important system which we will consider in this paper as the representative of the chiral systems, whether for pure physical treatment or technological applications.

\begin{figure}[!t]
\centering
\includegraphics[scale=0.4]{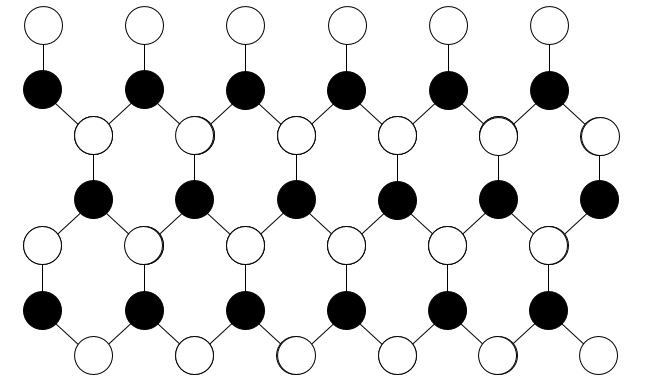}
\caption{Honeycomb lattice model. We can see the bipartite lattice by the different colors depicted, although they are all carbon.}
\label{figura1}
\end{figure}

The graphene is pictured in a honeycomb structure made of carbons in each vertex, as depicted in FIG. \ref{figura1}, and can be quantified by the Hamiltonian (\ref{eq46}). This structure results in the chiral symmetry (\ref{eq47}).  Physically, the latter introduces the coexistence of a electron-hole pair in the Dirac Sea. In fact, it was used to study some important theoretical issues about the transport in graphene, such as Klein paradox\cite{ref16} and, for our concern, to study the charge conjugation symmetry. Here we intend to study the transport properties of graphene considering further the parity and time reversal symmetries. A very straightforward way to couple these symmetries together is through the scattering matrix formalism, which we will develop now.

Considering a generic system with $\alpha$ ideal leads coupled to a cavity. The leads will transport carries from the reservoirs to the cavity, and also from the cavity to the detectors. For this reason, the solution for the lead $\alpha$ can be written as
\begin{equation}\label{eq1}
\psi_\alpha (x,y) = \displaystyle\sum_{n = 1}^{N_\alpha}(a_n^{(\alpha)} \psi_n^{- (\alpha)} + b_n^{(\alpha)} \psi_n^{+(\alpha)}),
\end{equation}
where we consider the cavity (scattering center) as the reference point. The coefficients $b_n^{(\alpha)}$ and $a_n^{(\alpha)}$ are generically called the output and input amplitudes, respectively, while $N_\alpha$ is the number of open channels in the $\alpha$-th lead. The wave functions $\psi^{\pm(\alpha)}_n$ are plane waves since we are assuming very far detectors (and emitters) from the cavity. We can represent the scattering picture of (\ref{eq1}) assuming that the output amplitudes result from an unitary transformation of the input amplitudes,
\begin{equation}\label{eq2}
B=SA,
\end{equation}
where $A$ and $B$ stand for the column vector of input and output amplitudes, respectively and $S$ is the scattering matrix which provides the relation between these amplitudes. We set the $S$ matrix as unitary, ($i.e.$, $SS^{\dagger}=\mathbb{I}$) in order to preserve the probability current. The scattering matrix fulfills all constraints because it accounts for all possible interactions that affect the transport properties of the system.
Since we are treating the graphene case, the scattering matrix must fulfill another constraint due to chirality, which is\cite{ref11}
\begin{equation}\label{eq1'}
S=\Sigma_z S^\dagger\Sigma_z.
\end{equation}
\begin{figure}[!t]
\centering
\includegraphics[scale=0.1]{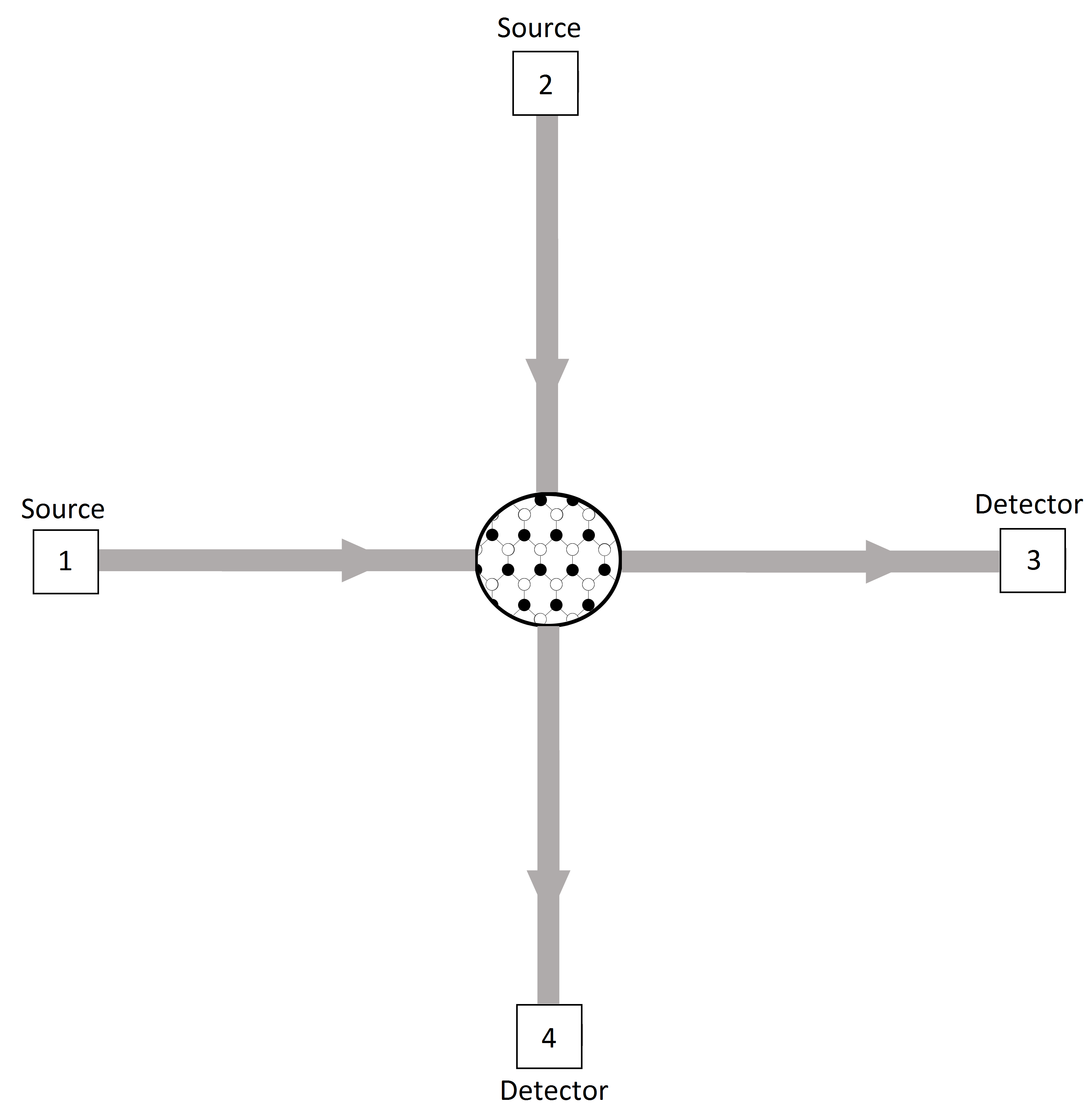}
\caption{Graphene as the scattering center transmitting two particles of different leads.}
\label{figura2}
\end{figure}
We can  establish (\ref{eq1'}) from (\ref{eq47}) adopting a Hamiltonian-dependent model to the scattering matrix ({\it e.g.} Mahaux-Weidenm\"{u}ller scattering model\cite{ref17}), we recover $S$ by the substitution $S(H,E)\rightarrow S(-\Sigma_z H \Sigma_z,-E)$. We turn our attention to the quantitative aspect of the scattering problem.

Usually, one may write the scattering matrix' entries as
\begin{equation} \label{ScM}
S =
\left( \begin {array}{cc} \mathbf{r}&\mathbf{t}^\prime\\ \noalign{\medskip}\mathbf{t}&\mathbf{r}^\prime\end {array}
 \right),
\end{equation}
with $\mathbf{r}$, $\mathbf{r}^\prime$ and $\mathbf{t}$, $\mathbf{t}^\prime$ representing the reflection and transmission blocks of $S$, respectively. We can apply the scattering formalism to the simplest graphene setup, consisting of two incident particles in different leads, each one with a single open channel. The graphene can be viewed as a scattering center which allows the particles to be transmitted or reflected to detector leads, as depicted in the Fig. \ref{figura2}.

Once we establish the conditions that reproduce the graphene setup, we can use (\ref{eq2}) to write its scattering equation according to Fig. \ref{figura2},
\begin{eqnarray} \label{eq3}
&\left( \begin{array}{c}
b_1\\
b_2\\
b_3\\
b_4
\end{array} \right)=e^{i\varphi}\left( \begin{array}{cccc}
0 & 0 & r^* & t^* \\
0 & 0 & t^* & r^* \\
r& t & 0 &0\\
t & r &0 &0
\end{array} \right)\left( \begin{array}{c}
a_1\\
a_2\\
a_3\\
a_4
\end{array} \right),&\nonumber\\
&\left( \begin{array}{c}
b_1\\
b_2\\
b_3\\
b_4
\end{array} \right)=S_\textrm{gph}\left( \begin{array}{c}
a_1\\
a_2\\
a_3\\
a_4
\end{array} \right),&
\end{eqnarray}
which $r$ and $t$ express the reflection and transmission amplitudes of the mirror, respectively. The phase $\varphi$, evaluated to $[0,\pi/2]$, can be viewed as a degree of chirality. The maximum chirality is achieved setting $\varphi = \pi/2$ when (\ref{eq1'}) is fulfilled, physically the phase $\varphi$ is the presence of impurities or doping in the graphene's structure\cite{ref25}. For bosons transported through the cavity, this quantity plays the same role as Y. Tang and A. Cohen defined Optical Chirality\cite{ref27}. For $\varphi=0$, we have the Hanbury Brown-Twiss interferometer(HBT)\cite{ref26}. The 4x4 scattering matrix in (\ref{eq3}) takes into account the backscattering in lead 3 and 4, resulting in non-trivial values to $b_1$ and $b_2$ which will emerge when we introduce barriers in each lead. Also another interesting fact we observe is that $S_\textrm{gph}$ obeys the chiral symmetry condition, (\ref{eq1'}). For our conceptual proposal, without loss of generality, we set the reflection and transmission probabilities of the cavity as equivalent, $i.e.$, $-r=it=\sqrt{2}/2$. In order to find the probabilities to detect both particles in lead $3$ or $4$ ($P(33)$ and $P(44)$, respectively), and one particle in lead $3$ and $4$ ($P(34)$), we consider each probability amplitude in (\ref{eq3}) as field operators acting in the vacuum state. Observing the setup depicted in Figure \ref{figura2}, the input state can be written as the creation operators $\hat{a}_1^\dagger$ and $\hat{a}_2^\dagger$ in the vacuum state. Then, one can calculate the probability to detect both particles in the same, or different, arms taking the square modulus of the projection of such input state and the respective state of interest, {\it i.e.}, $P(ij)=|\braket{0|\hat{b}_i \hat{b}_j \hat{a}_1^\dagger \hat{a}_2^\dagger |0)}|^2$. Using (\ref{eq3}) and the commutation relations for fermions, or bosons, the probabilities of detection in the output arms are given by
\begin{eqnarray}
&\textrm{P}(33)=\textrm{P}(44)= \dfrac{1}{4}(1-\epsilon |I|^2)& \label{eq4},\\
&\textrm{P}(34)= \dfrac{1}{2}(1+\epsilon |I|^2)\label{eq5}.&
\end{eqnarray}
Eqs.(\ref{eq4},\ref{eq5}) are valid for both fermions and bosons and the parameter $\epsilon$ inform two possibles algebras used: if $\epsilon=1$, we have the fermionic case, and, if $\epsilon=-1$, we have the bosonic's. Furthermore, $I$ is the overlap between the incident particles states which correlates with each other due to their indistinguishability, $I$ informs the simultaneity which the particles were emitted into the system: if $I=1$, the particles are emitted exactly at the same time, whereas $I=0$, the particles are emitted with a sufficiently large time delay. Despite the present system is the simplest case of a graphene setup, in the perfect overlapped situation, it manifests the very known Hong-Ou-Mandel(HOM) effect\cite{ref18}: to the bosonic case, we observe a bunching behavior of the carriers, to fermions, anti-bunching. The latter is known as eletronic HOM effect\cite{ref19,ref20}.

\begin{figure}[!t]
\centering
\includegraphics[scale=0.13]{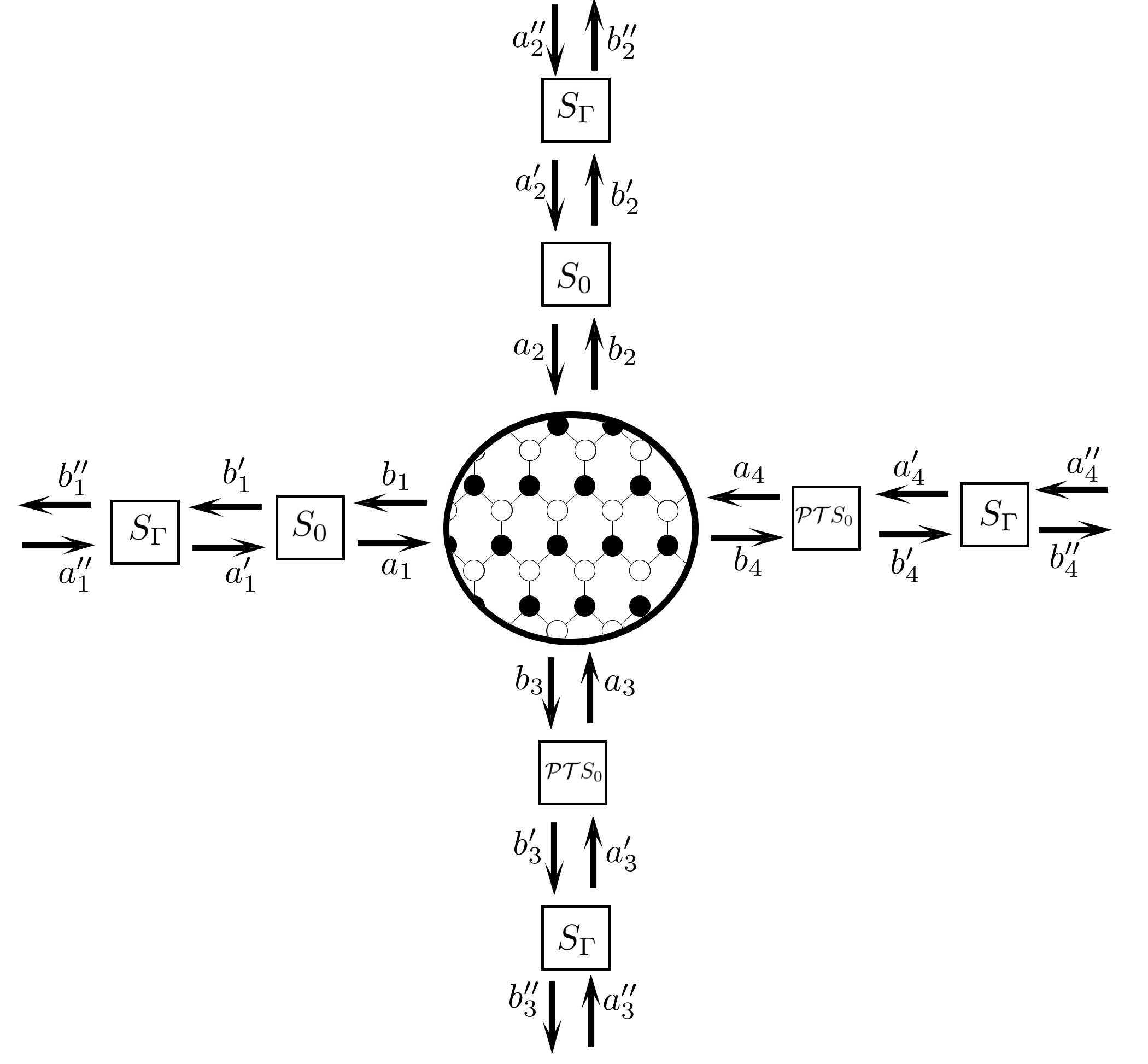}
\caption{Amplifying-Absorber sections and barriers coupling to the graphene setup\cite{ref3}.}
\label{figura3}
\end{figure}

Now we start the formal analysis to deduce the above system with PT symmetry.  In order to explore the PT-symmetry in the previous experiment, we need to implement a amplifying-absorber mechanism \cite{ref2} in order to find means to break parity and time reversal symmetries, thus, performing an extension of the one dimensional case \cite{ref3}, we couple two amplifying sections, in leads $1$ and $2$, with the scattering matrix (Fig.\ref{figura3})
\begin{eqnarray} \label{eq6}
&\left( \begin{array}{c}
b^\prime_i\\
a_i
\end{array} \right)=\left( \begin{array}{cc}
0 & t_0 \\
t_0 & 0
\end{array} \right)\left( \begin{array}{c}
a^\prime_i\\
b_i
\end{array} \right),&\nonumber \\
&\left( \begin{array}{c}
b^\prime_i\\
a_i
\end{array} \right)=S_0\left( \begin{array}{c}
a^\prime_i\\
b_i
\end{array} \right),&
\end{eqnarray}
where $t_0$ is an experimental parameter which rules the amplifying rate section and $i=1,2$. In order to maintain the PT-symmetry, it is necessary to balance the opposite arm by coupling two absorber sections in leads $3$ and $4$. The scattering equation is given by the result of the parity and time reversal operators, $\mathcal{P}$ and $\mathcal{T}$, by acting on $S_0$. Following the same procedure developed in \cite{ref3} to determine the form of such operators in the one dimensional case, we have
\begin{eqnarray} \label{eq7}
&\left( \begin{array}{c}
b^\prime_{i+2}\\
a_{i+2}
\end{array} \right)=\mathcal{PT}S_0\left( \begin{array}{c}
a^\prime_{i+2}\\
b_{i+2}
\end{array} \right),&\nonumber \\
&\left( \begin{array}{c}
b^\prime_{i+2}\\
a_{i+2}
\end{array} \right)=\sigma_x (S^*_0)^{-1} \sigma_x\left( \begin{array}{c}
a^\prime_{i+2}\\
b_{i+2}
\end{array} \right),&\nonumber \\
&\left( \begin{array}{c}
b^\prime_{i+2}\\
a_{i+2}
\end{array} \right)=\left( \begin{array}{cc}
0 & \frac{1}{t_0^*} \\
\frac{1}{t_0^*} & 0
\end{array} \right)\left( \begin{array}{c}
a^\prime_{i+2}\\
b_{i+2}
\end{array} \right),&
\end{eqnarray}
where $\sigma_x$ is the Pauli matrix. Eq. (\ref{eq7}) is expected to hold since we adopted the solution (\ref{eq1}) to a resonator \cite{ref4}. We complete the coupling of the sections in graphene setup substituting $a$ and $b$ of (\ref{eq6}) and (\ref{eq7}) in the graphene scattering equation (\ref{eq3}), and we get
\begin{eqnarray}\label{S+Sc}
&\left( \begin{array}{c}
b^\prime_1\\
b^\prime_2\\
b^\prime_3\\
b^\prime_4
\end{array} \right)=-e^{i\varphi}\dfrac{\sqrt{2}}{2}\dfrac{t_0}{t_0^*}\left( \begin{array}{cccc}
0 & 0 & 1&-i \\
0 & 0 &-i& 1 \\
 1&  i& 0& 0\\
 i&  1& 0& 0
\end{array} \right)\left( \begin{array}{c}
a^\prime_1\\
a^\prime_2\\
a^\prime_3\\
a^\prime_4
\end{array} \right).&
\end{eqnarray}
We note that the scattering matrix in (\ref{S+Sc}) no longer reproduces the Chiral symmetry as in (\ref{eq3}), but (\ref{S+Sc}) still reproduces the same outcome of the standard graphene setup, (\ref{eq4}) and (\ref{eq5}), as will be seen in section IV. Although we do recover former symmetry fixing the strong condition to the sections Im$(t_0^2)=0$, hence the imaginary part of $t_0^2$ is responsible for the crossover between the cases with graphene and the graphene-PT. Still working on FIG.(\ref{figura3}), we couple a tunneling barrier in each arm, aiming to explore the resonant regime. The scattering equation of the barriers is
\begin{eqnarray} \label{eq8}
&\left( \begin{array}{c}
b^{\prime\prime}_j\\
a^\prime_j
\end{array} \right)=-\left( \begin{array}{cc}
\sqrt{1-\Gamma} & i\sqrt{\Gamma} \\
i\sqrt{\Gamma} & \sqrt{1-\Gamma}
\end{array} \right)\left( \begin{array}{c}
a^{\prime\prime}_j\\
b^\prime_j
\end{array} \right),&\nonumber \\
&\left( \begin{array}{c}
b^{\prime\prime}_j\\
a^\prime_j
\end{array} \right)=S_\Gamma\left( \begin{array}{c}
a^{\prime\prime}_j\\
b^\prime_j
\end{array} \right),&
\end{eqnarray}
where $\Gamma$ is the barrier's transmission probability and $j=1,2,3,4$. Proceeding as in the previous case, we substitute $a^\prime$ and $b^\prime$ in (\ref{S+Sc}), and after some algebra we find the scattering matrix
\begin{equation}\label{eq9}
S_\textrm{T} =  \dfrac{1}{1-e^{2i\left(\varphi+2\theta\right)}(1-\Gamma)}\left( \begin{array}{cccc}
s & 0 & s^\prime & -is^\prime \\
0 & s & -is^\prime & s^\prime \\
s^\prime & is^\prime & s & 0 \\
is^\prime & s^\prime & 0 & s
\end{array} \right),
\end{equation}
where we defined
\begin{eqnarray}
&s = \sqrt{1-\Gamma}\left[e^{2i\left(\varphi+2\theta\right)}-1\right] ,& \label{eq10}\\
&s^\prime =  e^{i\left(\varphi+2\theta\right)}\Gamma.\label{eq11}&
\end{eqnarray}
where, we have made the complex variable substitution $t_0=|t_0|e^{i\theta}$. Eq. (\ref{eq9}) represents the scattering matrix which combines the graphene scattering experiment with parity and time reversal symmetries. One may argue that the barriers would affect the symmetry in each lead of (\ref{S+Sc}), however it is not an issue since $S_\Gamma$ is invariant under $\mathcal{PT}$ transformation; $S_\Gamma=\mathcal{PT}S_\Gamma$, as one can verify. Concerning  the strong condition of parity and time reversal, the system will not be affected by such quantities when $\theta=n\pi/2$, for $n=0,1,...,$. This can be viewed as a special case where no phase is gained due to the amplification or attenuation due to the sections. Such interpretation will aid us to study the resonances of the above system. Also, we recover $S_\textrm{gph}$ in Eq.(\ref{eq3}) when we fix such values of $\theta$ and $\Gamma\rightarrow 1$, as expected, since there are no backscattering effects.

Although we have represented the system in a fairly simple form through the scattering matrix of the Eq.(\ref{eq9}), some techniques are very cumbersome to be used to determine its spectrum. One method would be immerse the system in a heat bath \cite{ref3, ref5, ref6}, and evaluate the response of the system by the noise due to the coupling bath-system using the fluctuation-dissipation theorem \cite{ref7} in order to, finally, analyze the attenuation of the output amplitudes in the detection leads \cite{ref8} where it will detect a resonance. Clearly this method is very hard to implement in the four leads detection apparatus. Henceforth, we can approach such issue by the noise functions using the scattering formalism.
\section{III. The Noise Functions.}

We begin the description of the formalism considering the whole system composed by reservoirs, leads and the cavity which is under investigation. The reservoirs will provide the particles to be transported through the system and its statistics, given by the Fermi-Dirac distribution for the fermionic case, or by the Bose-Einstein distribution, for the bosonic's. The leads are responsible for the propagation of the emitted particles to the cavity.  We can solve the two-dimensional Schr\"{o}dinger equation in the leads considering the $x$-axis the direction of propagation and the $y$-axis as the transversal modes, and represent the latter behavior by the single-particle quantum field operator given by
\begin{equation}\label{eq12}
\hat{\Psi}(\vec{r}_\alpha,t)= \displaystyle\sum_n^{N}\int_0^\infty \textrm{d}k_{ n} \psi(k_{n},x_\alpha)\chi_{n}(y_\alpha) \hat{a}_\alpha(k_{n})e^{-\frac{E(k_{ n})}{\hbar}t}
\end{equation}
$\chi_{\alpha n}$ are the transversal eingenfunctions, $\psi(k_{\alpha n},x)$ some function that concern all the propagating quantities of the $n$-th channel beside $\hat{a}_{\alpha n}$ which is the destruction operator which fulfill the standard fermionic($\epsilon=1$) and bosonic($\epsilon=-1$) algebra
\begin{equation}\label{eq18}
\hat{a}_{\alpha n} \hat{a}^\dagger_{\beta m} +\epsilon \hat{a}_{\beta m}\hat{a}^\dagger_{\alpha n}  = \delta_{\alpha \beta}\delta(m-n).
\end{equation}
For our purpose, it is necessary discard the temperature in this problem and reduce the problem to the low frequency limit. Although it is an oversimplification, it will be enough to explore the effect due the symmetries in the transport of particles. These assumptions will reflect in the noise description of the problem, physically, the properties of the system will be given only by the transmission of the particles and its overlap, or indistinguishability degree. It will reflect in the field operator (\ref{eq12}) as \cite{ref9}
\begin{equation}\label{eq13}
\hat{\Psi}(x_\alpha,t)= \int \textrm{d}k_{\alpha} \psi(k_{\alpha},x_\alpha)\hat{a}(k_{\alpha})e^{-i\frac{E}{\hbar}t}.
\end{equation}
We observe that the field operator creates a one mode particle in the $\alpha$-th lead. The field operator of Eq.(\ref{eq13}) will act in Fock space, since we are interested in the multiplet case. As we settled the usual destruction field operator, now it is possible create an input state to interact with the quantum dot, as in \cite{ref1}. We inject a particle in lead 1 and 2 then we have the input state
\begin{eqnarray}\label{eq14}
\ket{\Psi}= \hat{\Psi}^\dagger(x_1,t)\hat{\Psi}^\dagger(x_2,t)\ket{0}
\end{eqnarray}
Since we intend to explore the properties by the noise, we have to represent an observable related to the particle's transport, in order to find the noise of the process, we recall for current operator given by
\begin{equation} \label{eq15}
\hat{I}_\alpha(t)= \int \textrm{d}x_\alpha \hat{T}\left[\Psi^\dagger(x_\alpha,t)\Psi(x_\alpha,t)\right],
\end{equation}
where $\hat{T}$ is the normal ordering operator. As we are dealing only with processes of transmission of the particles after interact with the quantum dot (\ref{eq9}), one can determine the possible outcomes through the calculation of the correlation function, given by\cite{ref21}
\begin{eqnarray}\label{eq16}
&S_{\alpha \beta}(t) = \dfrac{1}{2} \braket{\Delta \hat{I}_\alpha \Delta \hat{I}_\beta+\Delta \hat{I}_\beta \Delta \hat{I}_\alpha }&
\end{eqnarray}
where $\Delta \hat{I}_\alpha = \hat{I}_\alpha-\braket{\hat{I}_\alpha}$ is the fluctuation of the current operator at the lead $\alpha$ and the state that evaluate $S_{\alpha\beta}(t)$ and $\Delta I_\alpha$ is given by (\ref{eq14}). With the aforementioned assumptions, we can use the Wiener-Khinchin theorem in (\ref{eq16}) to obtain a fairly simple expression to the spectral density at low frequencies \cite{ref22}
\begin{eqnarray}\label{eq17}
&S_{\alpha \beta}(\omega) = \braket{\Delta I_\alpha \Delta I_\beta}_\omega,&
\end{eqnarray}
which has the same form of the probability functions $P(ij)$. In equation (\ref{eq17}), $I_\alpha$ is the current defined in the frequency domain
\begin{eqnarray}\label{eq17'}
I_\alpha(\omega) = \dfrac{1}{2\pi}\int I_\alpha (t)e^{-i\omega t}\textrm{d}t
\end{eqnarray}
The above analysis was necessary in order to identify the correlation function, {\it e.g.} (\ref{eq4}) and (\ref{eq5}), as a noise function. This may extend the concept of noise to scattering experiments. For the case of analyzing only the transmission of the carriers, one must impose the low frequency regime, then the correlation functions will be given by (\ref{eq17}), which we call shot noise, or Poissonian noise, whether the carrier is a fermion or a boson\cite{ref23,ref24}. As we are analyzing the transmission properties due to the system depicted in Figure \ref{figura3}, it is necessary develop a scattering formalism of such process to derive the respective noise functions, as we do next.

Since we already obtained the scattering matrix (\ref{eq9}) for 4 propagating leads with one channel each, we consider the scattering formalism with one channel per lead coupled to the quantum dot, then we write down the relation between of input and output amplitudes linked by the scattering matrix entries
\begin{eqnarray}\label{eq18'}
\hat{a}_{i}=\displaystyle \sum_{j=1}^4 s_{ij} \hat{b}_j
\end{eqnarray}
where $s_{ij}$ is the scattering matrix entries. In fact, we established (\ref{eq18'}) based on (\ref{eq2}). Then, with an analogous procedure used to derive (\ref{eq4}) and (\ref{eq5}), we project the input state (\ref{eq14}) on the outcome state of interest and take its modulus square, we achieve all the possible configurations of a 4x4 scattering matrix system given by
\begin{widetext}
\begin{eqnarray}
\label{eq19} &\big( \Delta \hat{I}_{(1,2,3,4)} \big)^2  = |s_{1(1,2,3,4)}|^2|s_{2(1,2,3,4)}|^2\left(1-\epsilon |I|^2\right), \\
\label{eq20} &\braket{\Delta \hat{I}_1 \Delta \hat{I}_{(2,3,4)}} = |s_{11}|^2|s_{2(2,3,4)}|^2 + |s_{1(2,3,4)}|^2|s_{21}|^2 -2\epsilon |I|^2\textrm{Re}\left(s_{11}s_{2(2,3,4)}s^*_{1(2,3,4)}s^*_{21} \right), \\
\label{eq21} &\braket{\Delta \hat{I}_2 \Delta \hat{I}_{(3,4)}} = |s_{12}|^2|s_{2(3,4)}|^2 + |s_{1(3,4)}|^2|s_{22}|^2 -2\epsilon |I|^2\textrm{Re}\left(s_{12}s_{2(3,4)}s^*_{1(3,4)}s^*_{22} \right),&\\
\label{eq22} &\braket{\Delta \hat{I}_3 \Delta \hat{I}_{4}} = |s_{13}|^2|s_{24}|^2 + |s_{14}|^2|s_{23}|^2 -2\epsilon |I|^2\textrm{Re}\left(s_{13}s_{24}s^*_{14}s^*_{23} \right).&
\end{eqnarray}
\end{widetext}
Eqs.(\ref{eq19})-(\ref{eq22}) allow us to determine all possible outcomes of any four terminal system, even with backscattering and any symmetry embedded. It is important to note that we no longer refer to the above correlation functions as probabilities, $P(ij)$, but as noise functions, $\braket{\Delta I_i \Delta I_j}$. Such equations immediately inform the shot noise of the system, as we are considering the system at zero temperature and low frequency. Finally, we extend our considerations to both fermionic and bosonic cases through the index $\epsilon$ and  also considered  the indistinguishability degree between the particles by the overlap integral
\begin{equation}\label{eq26}
I = \int \textrm{d} k\textrm{d}x \psi^*(x_a,k_a)\psi(x_b,k_b).
\end{equation}
It is important to notice that applying (\ref{eq3}) and (\ref{S+Sc}) in the shot noise equations, we immediately obtain the probabilities (\ref{eq4}) and (\ref{eq5}), and recover the HOM statistics, as expected. Another feature is that equation (\ref{eq19}) shows that it still preserves the exclusion principle for fermions, since we have a perfect indistinguishable pair, $I=1$, then the mean squared fluctuation of the current operator is zero.

Having previously derived the graphene-like cavity with PT symmetry scattering matrix, represented in Eq.(\ref{eq9}), now we can apply the scattering matrix elements in the shot noise functions, (\ref{eq19})-(\ref{eq22}), and obtain, after some algebra, the graphene with PT symmetry statistics
\begin{widetext}
\begin{eqnarray}
\label{eq36}&(\Delta \hat{I}_1)^2=(\Delta \hat{I}_2)^2=0,& \\
\label{eq37}&(\Delta \hat{I}_3)^2=(\Delta \hat{I}_4)^2= \dfrac{1}{4}\dfrac{\Gamma^4}{\Big [ 1+(1-\Gamma)^2-2(1-\Gamma)\cos(\varphi+2\theta)\Big ]^2}(1- \epsilon|I|^2),& \\
\label{eq38}&\Braket{\Delta \hat{I}_1\Delta \hat{I}_2}= \dfrac{4(1-\Gamma)^2\Big [ 1-\cos(\varphi+2\theta)\Big ]^2}{\Big [ 1+(1-\Gamma)^2-2(1-\Gamma)\cos(\varphi+2\theta)\Big ]^2},&\\
\label{eq39}&\Braket{\Delta \hat{I}_1\Delta \hat{I}_3}= \Braket{\Delta \hat{I}_2\Delta \hat{I}_4}=\Braket{\Delta \hat{I}_1\Delta \hat{I}_4}= \Braket{\Delta \hat{I}_2\Delta \hat{I}_3}=\dfrac{2\Gamma^2(1-\Gamma)\Big [ 1-\cos(\varphi+2\theta)\Big ]^2}{\Big [ 1+(1-\Gamma)^2-2(1-\Gamma)\cos(\varphi+2\theta)\Big ]^2} ,&\\
\label{eq40}&\Braket{\Delta \hat{I}_3 \Delta \hat{I}_4}=\dfrac{1}{2}\dfrac{\Gamma^4}{\Big [ 1+(1-\Gamma)^2-2(1-\Gamma)\cos(\varphi+2\theta)\Big ]^2}(1+ \epsilon|I|^2).&
\end{eqnarray}
\end{widetext}
If it is applied $\varphi=0$, one will obtain the statistics of a Hanbury Brown-Twiss apparatus embedded with parity and time reversal symmetry\cite{ref10}. Equations (\ref{eq37}) and (\ref{eq40}) are the noise due to the full transportation of the carriers. Equation (\ref{eq39}) represents the transportation of one carrier and the reflection of the other and  equation (\ref{eq38}) represents no transport at all. In the HBT regime, one can achieve the resonances setting the breaking PT symmetry condition to $\theta$: only the full transportation noise equations will be non-trivial. Also, we can analyze how the slightest doping will affect the resonant regime in the former case. Expanding the above equations to the next non-trivial order of $\varphi$, one gets immediately the corrections due to the backscattering caused by impurities effects through the noise,
\begin{eqnarray}\label{eq41}
&\braket{(\Delta I_{tr})^2} = f_\epsilon(I)\left[1+2\dfrac{1-\Gamma}{\Gamma^2}\varphi^2\right],&\\
&\braket{(\Delta I_{bs})^2}\sim \varphi^4,&
\end{eqnarray}
where $f_\epsilon(I)$ is relative to the overlap term of the respective function, $\braket{(\Delta I_{tr})^2}$ is related to the noise functions which describes the full transmission of the carriers, and $\braket{(\Delta I_{bs})^2}$, is the noise functions with any type of backscattering. It is interesting to note that in a resonant regime($\Gamma\rightarrow 0$) of a doped system border effects of the cavity increase as corrections of the noise functions go up. It is possible too make the same expansion around $\varphi=\pi/2$, which will produce the noise functions with corrections due to impurities in graphene.

Also, it is possible to find every resonant regime condition of a more general system, as easily seen in the above equations when $\cos(\varphi +2\theta)=1$, it will be obtained through a relation between the PT-symmetry parameter and the pureness degree of such a system
\begin{equation}\label{eq43}
\varphi+2\theta = 2 \pi n
\end{equation}
Eq. (\ref{eq43}) shows us that the true responsible for a full transportation of the particles by the system is how the PT-symmetry relates itself with the intrinsic symmetry, since the noise functions turn to be independent of transmission barriers and the nature of the carriers. The generality of this statement is not casual: if one is interested in another Cartan class, it will be as possible to find the resonance condition by a similar treatment addressed here. 

\begin{figure*}[t]
\begin{minipage}[t]{0.48\linewidth}
\subfigure[$(\Delta \hat{I}_1)^2=(\Delta \hat{I}_2)^2$ for fermions and $\braket{\hat{I}_3\hat{I}_4}$ for bosons.]{\includegraphics[scale=0.6]{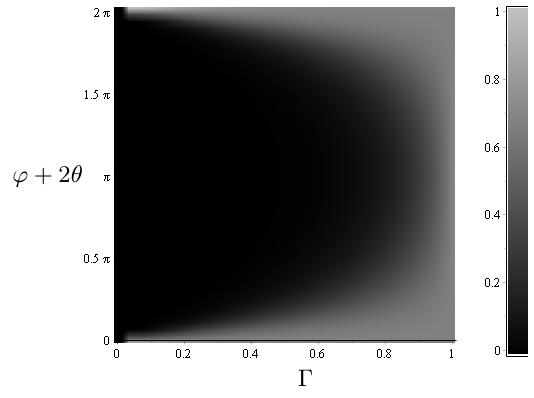} }\\
\subfigure[$\braket{\hat{I}_1\hat{I}_2}$.]{\includegraphics[scale=0.6]{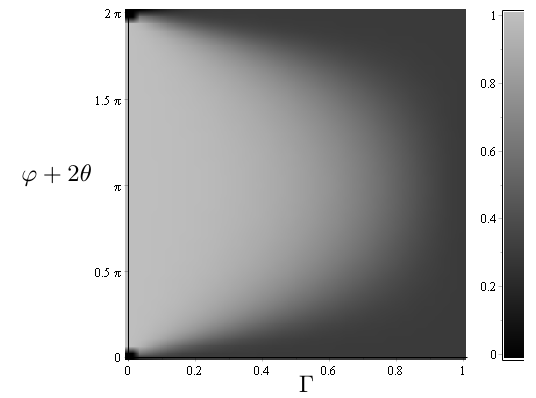}}
\end{minipage}
\begin{minipage}[t]{0.48\linewidth}
\subfigure[$\braket{\hat{I}_{[1,2]}\hat{I}_{[3,4]}}$.]{ \includegraphics[scale=0.6]{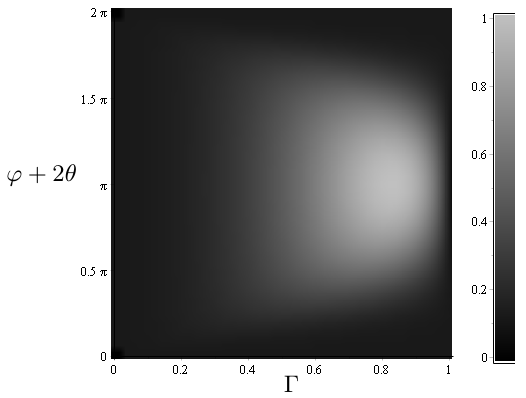}} \\
\subfigure[$\braket{\hat{I}_3\hat{I}_4}$ for fermions and $(\Delta \hat{I}_1)^2=(\Delta \hat{I}_2)^2$ for bosons.]{ \includegraphics[scale=0.6]{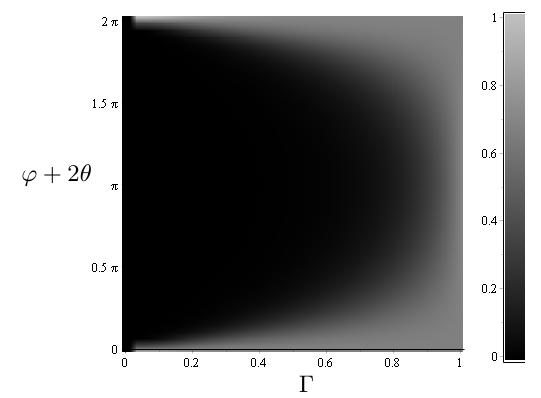}}
\end{minipage}
\caption{The noise functions (\ref{eq36})-(\ref{eq40}). Apart from the extreme values 0 and $2\pi$, the system become strongly dependent on $\Gamma$. } \label{figura4}
\end{figure*}

Moreover, it is possible to obtain more possible physical situations beside (\ref{eq43}), as depicted in FIG. \ref{figura4}. As we can notice, in order to have non-trivial values to the backscattering noise equations, FIG. \ref{figura4} (b) and (c), one must take into account the values of the transmission barriers. When the symmetry parameters do not fulfill (\ref{eq43}), one can identify transitions between the backscattering to the full transmission case, and vice-versa, by varying $\Gamma$.

From a practical point of view, the studying of the relation between chirality and PT symmetry will reduce the values of $\varphi$ to its extremes(0 or $\pi/2$) and it will given different quantization conditions when one reaches the closed cavity. Analyzing the perfect chirality condition, $\varphi=\pi/2$, and the resulting noise functions, we reach to the new quantization condition to a graphene-like system through the Parity and Time Reversal breaking
\begin{eqnarray}\label{eq42}
\theta_{gph} = n\pi-\dfrac{\pi}{4}.
\end{eqnarray}
The result (\ref{eq42}) gives the entire resonances points of a ${\cal \chi/PT}$ symmetric system, provided the existence of such constraint between these two symmetries, the $\chi$ and the PT. Also, we  look at the direct effect of the PT symmetry on an observable such as the shot noise and how a non-Hermitian Hamiltonian formulation can assist to find the energies of a system at any amount of doping. 

There are some interesting features in graphene experiment when we compare with the HBT and the HBT-PT case. As was studied in \cite{ref10}, the usual HBT scattering matrix possess the Hermitian condition, following the basic postulates of quantum mechanics. On other hand, the HBT-${\cal PT}$ system do not possess the latter condition, but does have the parity and time reversal symmetry. When one adjusts the HBT-{\cal PT} sections in order to obey the condition Im$(t_0^2)=0$, the usual HBT scattering matrix and correlation functions are obtained, even in the presence of the barriers, which indicates when the {\cal PT} symmetry of the system is broken. The situation is fairly different when we substitute the HBT system by a graphene. We learned above that graphene has the chiral symmetry expressed through the constraint $S = \Sigma_z S^\dagger \Sigma_z$, which causes the losing of hermiticity condition of its scattering matrix, even though the correlation functions are HBT-kind. 
 Then, coupling the sections and barriers, we lose the chiral condition and gain parity and time reversal symmetry, however when the sections obey Im$(t_0^2)=0$, the system enters in resonance regime and turn again chiral symmetric. Apparently, both HBT-${\cal PT}$ and graphene systems share the property of recovering their initial condition, Hermiticity and chirality respectively, when we set Im$(t_0^2)=0$, but fundamentally they differ in its noise functions. The former do not explicitly depends of the barriers when we take away PT symmetry, this is clearly not the case of graphene, which, in the same amplifying-attenuation regime, depends directly of the tunnel barriers, as we can see substituting in the noise equations the values $\varphi=\pi/2$ and $\theta=n\pi/2$, the latter being the resonance condition to the HBT system.

\section{IV. Conclusion}

In the present paper, we have analyzed a $\chi$/PT symmetric system, such as graphene where chirality refers to the presence of a pair electron-hole, and encountered its resonances through a simple mechanism of setting the proper values of the quantities related to the symmetry of such system. Furthermore, it was found that the correction was due to the presence of impurity in a HBT-like system. For experimental purposes, one can set an experimental device symmetric by parity and time reversal following the procedure pointed in \cite{ref10}, and may encounter its resonances by the present procedure. Moreover, it is important to remark that one can find the resonance regime of any system, given its scattering matrix, just doing the same analysis of the symmetry parameters, $\varphi$ and $\theta$. Its possible either to construct further formulations of scattering matrices, with other symmetry parameters, since the proper constraint of the Cartan classes is fulfilled. Besides, the formalism used here leans toward the profound discussion of a more fundamental formulation of quantum mechanics based in symmetry principles. Although we did not discussed its proper implications, one can find some insightful discussions in references \cite{ref28, ref29, ref30, ref36, ref37}. Finally, we envisage that the
${\cal \chi/PT}$ symmetry may find applications in microbiology and the DNA research in general\\

\section*{Acknowledgments}

This work was partly supported by the  the Brazilian agencies, CNPq, CAPES, FACEPE and FAPESP. MSH acknowledges a Senior Visiting Professorship granted by CAPES/ITA. This work is also supported by the project INCT-FNA Proc. No. 464898/2014-5.

\end{document}